# NGC 1912 and NGC 1907: a close encounter between open clusters ?

M.R. de Oliveira[1,2], A. Fausti[2], E. Bica[2], and H. Dottori[2]

[1] LERMA, Observatoire de Paris, 61 Av. de l'Observatoire, F-75014 Paris, France
[2] Instituto de Fisica-UFRGS, CP 15051, CEP 91501-970 POA - RS, Brazil

**Abstract.** The possible physical relation between the closely projected open clusters NGC 1912 (M 38) and NGC 1907 is investigated. Previous studies suggested a physical pair based on similar distances, and the present study explores in more detail the possible interaction. Spatial velocities are derived from available radial velocities and proper motions, and the past orbital motions of the clusters are retrieved in a Galactic potential model. Detailed N-body simulations of their approach suggest that the clusters were born in different regions of the Galaxy and presently experience a fly-by.

**Table 1.** Basic data for NGC 1912 and NGC 1907. (1) Cluster name; (2) J2000 $\alpha$; (3) J2000 $\delta$; (4) age (Myr); (5) reddening; (6) projected separation; (7) distance to the Galactic centre.

| Name | $\alpha$ | $\delta$ | Age | E(B-V) | separ. | $R_g$ |
| | h : m : s | $^o$ : ' : " | | | pc | kpc |
| (1) | (2) | (3) | (4) | (5) | (6) | (7) |
| NGC1907 | 05:28:06 | 35:19:30 | 400 | 0.40 | 18.8 | 9.8 |
| NGC1912 | 05:28:43 | 35:51:18 | 250 | 0.23 | | 9.8 |

## 1. Introduction

In contrast to simple projection effects, star cluster physical interactions can be basically classified as: (i) binary physical systems with common origin in a star-forming complex, necessarily with comparable age; (ii) binary physical systems arising from a spatial encounter with capture, which are expected to have different ages in most cases; (iii) a spatial encounter without capture.

The existence of star cluster pairs in the Magellanic Clouds have been investigated extensively by several authors in the last decade (Bhatia & Hatzidimitriou 1988, Bica et al. 1992, Vallenari et al. 1998, Dieball & Grebel 2000 and de Oliveira et al. 2000b). However, the identification of physical star cluster pairs in the Galaxy, with the exception of $h + \chi$ Persei, has not been an easy task owing to the strong possibility of projection effects. As we are seeing the Galaxy from inside the cluster distances should be taken into account. A list of candidates for physical cluster pairs in the Galaxy was proposed by Subramaniam et al. (1995). With the use of open cluster catalogues they could identify 18 probable cluster pairs with spatial separations less than 20 pc and suggested that about 8% of their data could be physical systems. Sub-

*Send offprint requests to*: marcio.oliveira@obspm.fr

ramaniam & Sagar (1999), with the use of CCD photometry, obtained ages, distances and reddening for two possible Galactic open cluster pairs and suggested that the pair NGC 1912/NGC 1907 was a good candidate for a physical pair. They advanced this possibility based on the resulting similar age values.

NGC 1907 is a cluster projected near NGC 1912 (M 38) with a projected separation of 18.8 pc (Subramaniam & Sagar 1999, hereafter SS99). Hoag (1966) obtained a distance of 1380 pc and a reddening of E(B-V)= 0.38 for NGC 1907. More recently, SS99 derived $d_\odot$=1785 ± 260 pc and E(B-V)=0.40. For NGC 1912, Hoag & Applequist (1965)Km obtained, with the use of photoelectric photometry, $d_\odot$= 870 pc and E(B-V)=0.27, while SS99 obtained $d_\odot$=1810 pc ± 265 pc and E(B-V)= 0.23. Reddening differences probably arise from the fact that the clusters are in the background of the Taurus-Auriga dust complex, where extinction is highly variable. By means of isochrone fitting SS99 derived ages of 400 and 250 Myr for NGC 1907 and NGC 1912, respectively. Age differences in this range are difficult to assert, within uncertainties. Taken at face values this difference is only marginally consistent with the hypothesis of a common origin in the same star-forming complex (Fujimoto and Kumai 1997). In Table 1 we show some information for this cluster pair as compiled by SS99.

In the present study we estimate the spatial velocities from available radial velocities and proper motions. The bundle of orbits allowed by the error bar in spatial velocity was retrieved by mean of N-body simulations in a

**Table 2.** Input parameters for the simulations. (1) Model; (2) number of particles of the cluster; (3) cluster total mass; (4) half-mass radius; (5) maximum radius of the cluster; (6) mean velocity of the particles (modulus); (7) softening parameter.

| Model | $N_{\rm part}$ | $M_{\rm T}$ $(M_\odot)$ | $R_{\rm h}$ (pc) | $R_{\rm max}$ (pc) | $V_{\rm md}$ (km/s) | $\epsilon$ (pc) |
|---|---|---|---|---|---|---|
| M1907 | 3072 | 3000 | 2.0 | 4.0 | 1.5 | 0.03 |
| M1912 | 4096 | 4200 | 3.1 | 6.0 | 1.5 | 0.03 |

Galactic potential model over 200 Myr. The simulation time is comparable to the ages and relaxation times of the clusters.

In Sect. 2 we present the initial conditions of the simulations. In Sect. 3 we present the results and discussion. Finally, the concluding remarks are given in Sect. 4. The present simulations are part of a series of N-body simulations to be fully explored in a coming paper.

## 2. The method and initial conditions

The N-body code used is the TREECODE algorithm (Hernquist 1987, Hernquist & Katz, 1989) where the force calculation between two particles is done by a softened Keplerian potential. We performed the simulations in a CRAY-T94 computer of the CESUP Laboratory of the Universidade Federal do Rio Grande do Sul. The units used in the simulation are pc, km/s, Myr and G=1.

The models are represented by a density distribution that follows a Plummer truncated polytrope (Aarseth et al. 1974) and have been generated in the same way as in de Oliveira et al. (1998) - hereafter ODB98. Cluster model dimensions were based on estimates for the real clusters (Lyngå 1987). The cluster masses were estimated from the best statistics absolute magnitude ($M_{\rm V}$) interval in common with the Hyades (-0.3 < $M_{\rm V}$ < 1.2), using colour magnitude diagrams (CMD) from SS99 and the WEBDA database (Mermilliod 1996), in the web page obswww.unige.ch/webda. The masses were estimated by scaling the number of members in the fiducial $M_{\rm V}$ interval, assuming comparable mass functions. Adopting for the Hyades a total mass of $\simeq 1000 M_\odot$ (Perryman et al. 1998), we obtained masses of $\simeq 3400 M_\odot$ and $2500 M_\odot$ for NGC 1912 and NGC 1907, respectively. In contrast to the equal mass models of ODB98 we introduced a mass spectrum that follows the Salpeter mass function, i.e. $dN/dm \propto m^{-2.35}$, and adopted a mass spectrum ranging from 0.5 to 3.0 $M_\odot$ for NGC1912 and 0.5 to 2.4 $M_\odot$ for NGC1907 according to the turnoff mass and age relation of Renzini & Buzzoni (1986). As we search here for a possible cluster encounter that occurred in a recent epoch, we simulated our models in time scales of the order of the relaxation time $t_{\rm rx}$ of the clusters, thus avoiding complications of long-term effects like core collapse and gravothermal oscillations which occur at $\simeq 15\, t_{\rm rx}$ (Cohn 1980, Makino 1996). The present objective is trying to detect any tidal effect in the outskirts of the clusters, and thus we neglected the effects of primordial binaries as they have little effect on the overall evaporation rate of the clusters (McMillan & Hut 1994).

As we are simulating the evolution of the clusters since T= -200 Myr we have to take into account the mass loss due to stellar evolution. To do this we calculated how the Salpeter mass function evolves back in time. Approximately 10% of the mass is lost for both clusters in the time interval T= -200 Myr (initial condition) to T= 0 (present condition). Also, we calculated the mass loss due to the dynamical evolution of the cluster in the presence of an external gratitational field, which causes evaporation of cluster stars. We assumed a typical value of mass loss by evaporation of 3% per relaxation time, according to Gnedin & Ostriker (1997). During the simulation the stellar evolution has not been included since it is expected to be more important for the very early evolution of the clusters (Terlevich 1987). The initial values for our models are presented in Table 2.

### 2.1. The Galaxy Model

The Galaxy potential was modeled by means of three components, disk, bulge and halo, combined to reproduce the rotation curve compiled by Allen & Martos (1986).

The disk model is that of Kuzmin & Kutuzov (1962). The Hernquist & Katz's (1989) TREECODE works with a triaxial generalization of the Kuzmin-Kutuzov model. The potential for this model is

$$\Phi_{\rm d}(w,z) = -\frac{GM_d}{(w^2 + z^2 + a^2 + b^2 + 2\sqrt{a^2 b^2 + b^2 w^2 + a^2 z^2})^{\frac{1}{2}}}$$

where $w$ and $z$ are the radius and height in cylindrical coordinates. $M_d$ is the total mass of the disk component, $a$ is the length scale in $w$ and $b$ is the scale in $z$.

The bulge model (Hernquist 1990) has the following potential expression:

$$\Phi_{\rm b}(r) = -\frac{GM}{r+c}$$

where $r = \sqrt{w^2 + z^2}$ is the radius in spherical coordinates and $c$ the length scale.

The halo was modeled with a logarithmic potential (Binney & Tremaine 1987),

$$\Phi_{\rm h}(w,z) = \frac{1}{2}v_0^2 ln(d^2 + w^2 + \frac{z^2}{q_\Phi^2}) + constant$$

where $v_0$ is the asymptotic halo velocity, $d$ is the length scale of the model and the *constant* is adjusted so that $\Phi_{\rm h}$ is zero at $r_{\rm c}$, the cutoff radius of the model. We assume a spherical halo ($q_\Phi = 1$). The parameters of the three components that fit the data, together with the data intervals used in the fits are listed in the Table 3.

**Table 3.** A non-linear curve fitting function was used to optimize the parameters of the galaxy model a, b, $M_d$, c, $M_b$ and d with precision 0.01. The data intervals are quoted in Col. 3.

| Parameters | Fit results | Constraints |
|---|---|---|
| Disk length scale in $w$: | $a = 4.61$ kpc | $(4000 < a < 7000)$ pc |
| Disk length scale in $z$: | $b = 0.40$ kpc | $(200 < b < 400)$ pc |
| Disk mass: | $M_d = 9.35 \times 10^{10} M_\odot$ | $(5.8 \times 10^{10} < M_d < 1.4 \times 10^{11}) M_\odot$ |
| Bulge length scale in $r$: | $c = 0.32$ kpc | $(300 < c < 800)$ pc |
| Bulge mass: | $M_b = 2.02 \times 10^{10} M_\odot$ | $(3 \times 10^9 < M_b < 4.6 \times 10^{10}) M_\odot$ |
| Halo asymptotic Velocity: | $v_0 = 190$ km/s | $(100 < v_0 < 250)$ km/s |
| Halo length scale: | $d = 14$ kpc | $(5000 < d < 15000)$ pc |

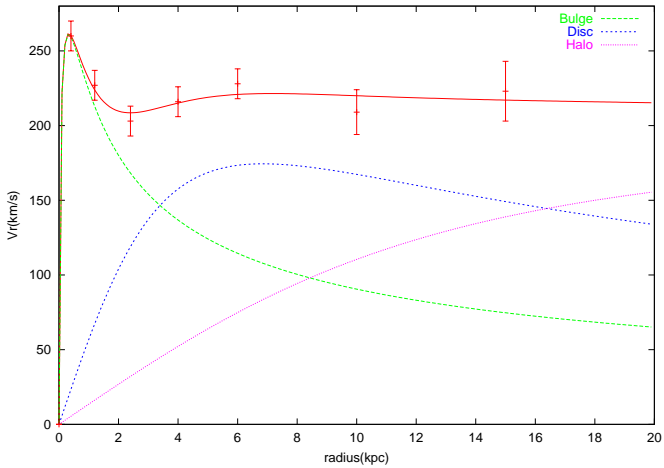

**Fig. 1.** Rotation curve resulting from the choice of the three mass components compared to the data points for the Galaxy, as compiled by Allen & Martos (1986).

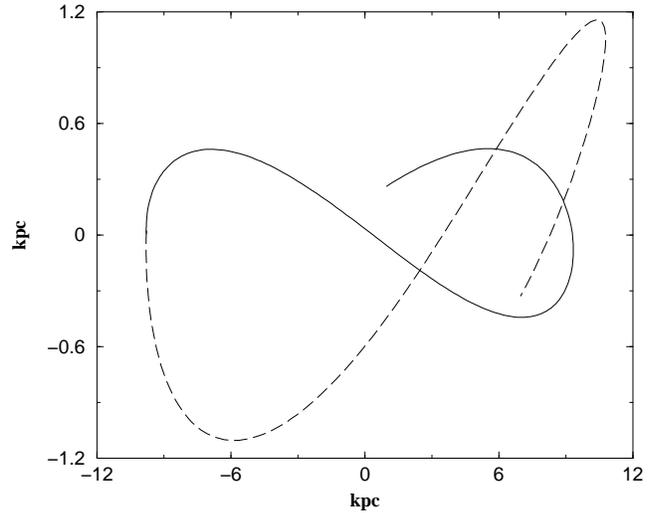

**Fig. 2.** Centre of mass orbit projected on the XZ plane for $-200 < T$ (Myr) $< 0$. Dashed line is N-body model M1912 representing the open cluster NGC 1912, and solid line is M1907 representing NGC 1907.

**Table 4.** Position and velocity components of the clusters in the Galactocentric inertial system for the present epoch.

| Object | X [pc] | Y [pc] | Z [pc] | V [km/s] | U [km/s] | W [km/s] |
|---|---|---|---|---|---|---|
| NGC1907 | 9783 | 231 | 9 | 6.16 | 212.29 | -19.74 |
| NGC1912 | 9781 | 242 | 22 | 10.90 | 213.87 | 41.97 |

The rotation curve resulting from the potential $\Phi = \Phi_d + \Phi_b + \Phi_h$ is plotted in Figure 1.

### 2.2. Orbital Initial Conditions

The kinematical initial conditions were derived from proper motions and radial velocities. Heliocentric radial velocities and mean proper motions have been obtained from the WEBDA open cluster database (Mermilliod 1996). For NGC 1907, the mean proper motions have been calculated from two sources, namely WEBDA and Tycho-2 Catalogue (Hog et al. 2000). From Tycho-2 we have considered five stars with magnitudes $V < 12.4$, which are not non-members according to WEBDA. We used the weighted average of these values to obtain the mean proper motions for NGC 1907. The radial velocities and available errors for NGC 1912 and NGC 1907 are $V_r = -1.00 \pm 0.58$ km/s and $V_r = 0.10 \pm 1.80$ km/s respectively. The proper motion values for NGC 1912 are $\mu_\alpha = 3.60 \pm 0.50$ mas/yr and $\mu_\delta = 1.90 \pm 0.50$ mas/yr, and for NGC 1907 are $\mu_\alpha = -0.81 \pm 0.73$ mas/yr and $\mu_\delta = -4.51 \pm 0.76$ mas/yr.

Using Johnson & Soderblom's (1987) method, we obtained the present day set of vector components $x_0$, $y_0$, $z_0$, $U_0$, $V_0$, $W_0$ (Table 4), which refer to a right-handed Cartesian Galactocentric inertial system where the Sun is at $X_\odot = -8$ kpc, $Y_\odot = 0$, $Z_\odot = 0$ and the z axis points towards the North Galactic Pole. In order to correct for the solar motion we used $U_\odot = 10$ km/s, $V_\odot = 240$ km/s, $W_\odot = 7.5$ km/s with $U_{LSR} = 0$, $V_{LSR} = 225$ km/s, $W_{LSR} = 0$.

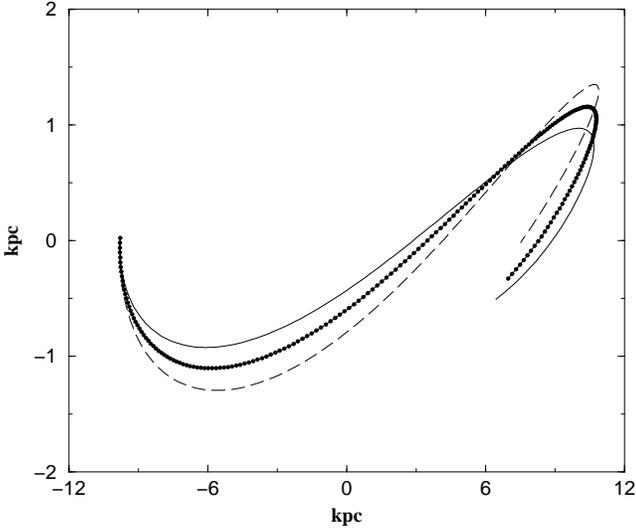

**Fig. 3.** Centre of mass orbit projected on the XZ plane for -200 < T (Myr) < 0 for three possible orbital models for NGC 1912. Dotted line is N-body model M1912 with the mean velocity values prsented in the text. The solid line is the model with velocity values corresponding to the error lower limits, while the dashed line corresponds to the upper limits.

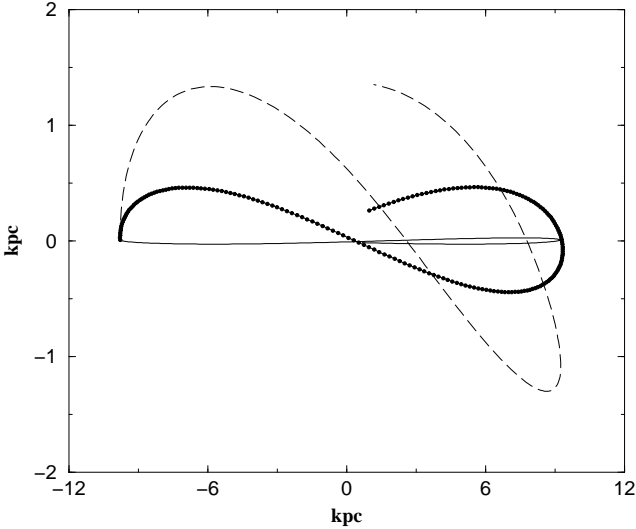

**Fig. 4.** Same as Fig. 3 for a set of possible orbit models for NGC 1907.

We left a point mass at $x_0$, $y_0$, $z_0$, $-U_0$, $-V_0$, $-W_0$, to evolve in the Galactic potential, to get $x_i$, $y_i$, $z_i$, $-U_i$, $-V_i$, $-W_i$ at an earlier time. Then, by reversing again the velocity components we obtained the initial conditions of our simulations $x_i$, $y_i$, $z_i$, $U_i$, $V_i$, $W_i$.

In Fig. 2 we plot the orbit of both clusters from T= -200 Myr up till now, according to the values obtained above.

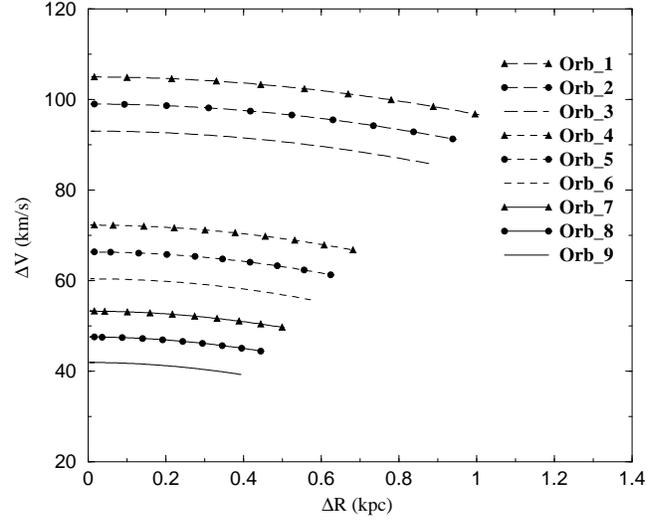

**Fig. 5.** Probing, within parameter uncertainties, other possible orbital solutions similar to that of Fig. 1. We plot the relative velocity against separation (modulus) for alternative models in the time interval T= -10 Myr to T= 0.

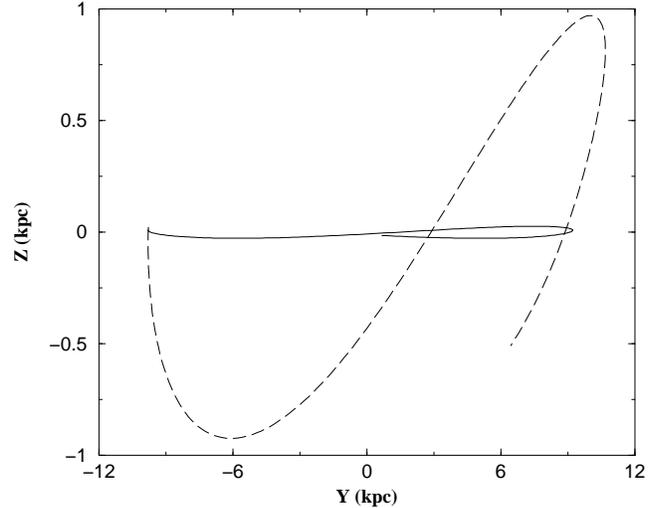

**Fig. 6.** XZ plane projection of closest approach model Orb-9 (Fig. 2) for -200 < T(Myr) < 0. Line symbols as in Fig. 2.

## 3. Results and Discussion

Around the orbital solution of Fig.2 we probed a grid of radial velocities and proper motion values within uncertainties in these quantities, trying to find out which one provided the smallest relative velocity and separation between the clusters. In Figs. 3 and 4 we show the orbital results for the NGC 1912 and NGC 1907 models. Each figure has three possible orbits, one corresponding to the mean values for the orbit in Fig. 2 and the other two to the extreme values derived by means of the upper and lower limits for the errors of the velocity components.

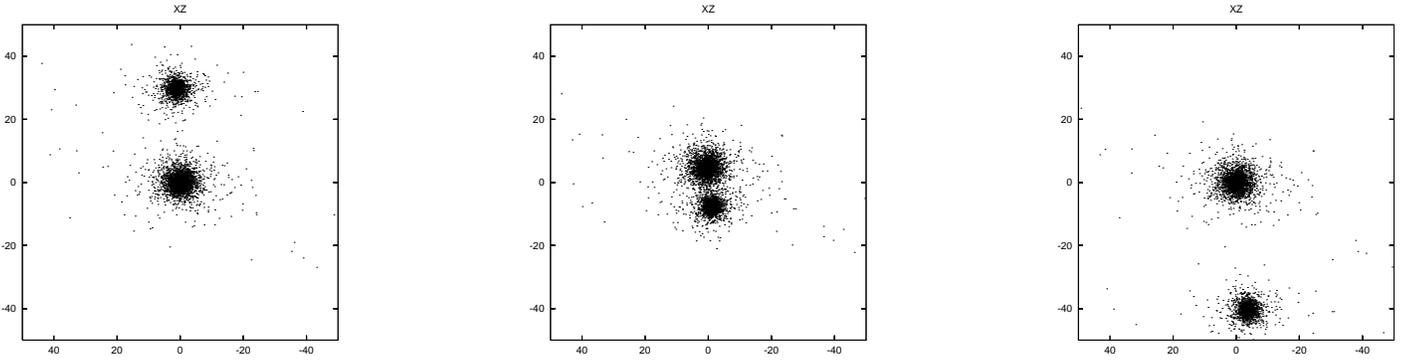

**Fig. 8.** XZ plane projection of the cluster encounter according to orbit model Orb-9 at three different times. From left to right: orbit model at T=-1.2 Myr, at present time (T=0) and in the near future (T = +0.8 Myr).

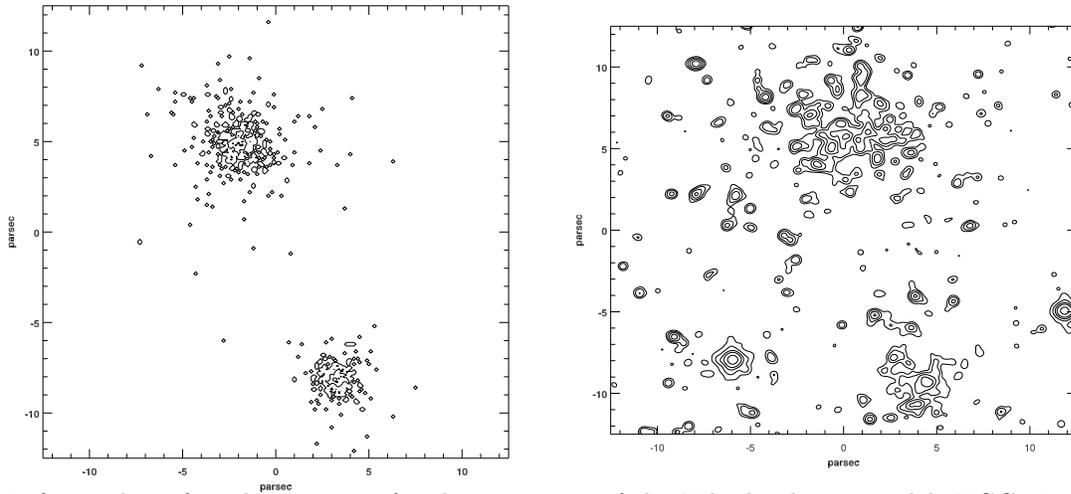

**Fig. 9.** Left panel: surface density map for the encounter of the N-body cluster models NGC 1912 (top) and NGC 1907 (bottom). Right panel: surface density map of the open clusters NGC 1912 (top) and NGC 1907 (bottom), as derived from the Digitized Sky Survey.

In Fig. 5 we show the relative velocity against separation (modulus) for all the possible orbital model combinations of Figs. 3 and 4 in the interval T= -10 Myr to T= 0. The best model is Orb-9 which gives a relative velocity $\Delta V$= 41.9 km/s for a centre to centre separation of 2.7 pc at T= -0.4 Myr. The orbits for both clusters are plotted in Fig. 6.

Since the radial velocity of NGC 1912 was based on one star (Sect. 2.1), we also tested other orbits including the measures of three more stars presented by Glushkova & Rastorguev (1981). Despite a high membership probability, these authors have classified them as non-members based on proper motions. In Fig. 7 we show the relative velocity against separation (modulus) for variations of the Orb-9 model taking into account the possibility of different values for the NGC 1912 radial velocity based on these additional stars. We conclude that the orbit is not much dependent on this uncertainty.

In Fig. 8 we show the encounter according to the Orb-9 orbit model for T= -1.2 Myr, T= 0 (present time) and T= +0.8 Myr. The possible orbit (Fig. 6) suggest that the clusters were formed in different parts of the Galaxy, which is corroborated by their differing ages (Fujimoto and Kumai 1997). We can also estimate, as first approximation, that the limiting velocity for a close orbit between the clusters is around one order of magnitude smaller than the relative velocity obtained for model Orb-9 at T= -0.4 Myr. The results suggest that a fast fly-by with transpassing occurred in the near past.

For isophotal analysis purposes we extracted digitized images of the pair NGC 1912/NGC 1907 from the DSS. The plate is a Palomar blue one. The PDS pixel values correspond to photographic density measures from the original plates, and are not calibrated. The method used here to obtain the isophotal contours is identical to de Oliveira et al. (2000a).

In Fig. 9 we compare the model clusters isodensity map on the sky at the present epoch with the NGC 1907 and NGC1912 isophotal contours.

Apparently there is no evidence of a bridge linking the clusters, neither for the N-body models nor for the open clusters themselves (Fig. 9). In a multi-mass system, due to mass segregation, low mass stars are preferentially stripped out by the external tidal galactic field, so one expect that tidal tails are mainly constituted by low mass stars (Combes et al. 1999). So, in principle, these tidal bridges, arising in isolated clusters, would not be easily detectable. However, in a cluster pair encounter we can expect mass loss by one of the members up to 50% of the initial cluster mass (Leon et al. 1999, de Oliveira et al. 2000a). Also, according to the study of possible interacting clusters in the LMC Leon et al. (1999) have calculated that the star upper limit mass of these tails are comparable to the most massive stars in these clusters. So we would expect that in a cluster encounter some massive stars are present in these tidal tails.

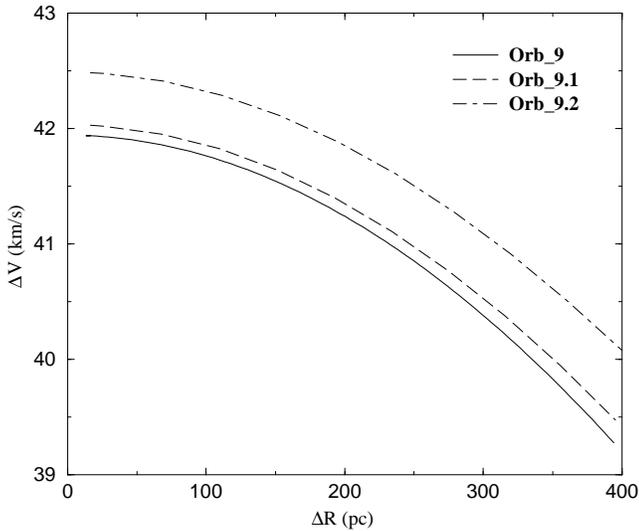

**Fig. 7.** Same as Fig. 5 for three different values for the radial velocity of model M1912: -1.0 (solid), -5.0 (dashed) and -9.0 km/s (dot-dashed) respectively.

The simulations allow us to check how intense a bridge would appear in a near capture relative velocity. In order to test such scenario we simulated a parabolic encounter reducing the relative velocity to 10% of the original value, which is beyond the error bar. In Fig. 10 we plot the on-the-sky model isodensity at T=0, after the closest approach. The simulation shows that a strong bridge arises in this case.

Captures would require to lower the relative velocity modulus in the last simulation which in turn would increases the bridge intensity. So, our model suggests that we are witnessing a pair fly-by.

## 4. Conclusions

In the present study we investigated if the closely projected open clusters NGC 1912 (M 38) and NGC 1907 are a binary cluster. We derived spatial velocities from available radial velocities and proper motions, and the past orbital motions of the clusters are retrieved in a Galactic potential model. According to our models, the clusters were formed in different parts of the Galaxy. The detailed N-body simulations of NGC 1912 and NGC 1907 encounter suggest a fly-by occurred in the near past. These simulations also shows that the faster the clusters approach the weaker the tidal debris in the bridge region, which explain why there is, apparently, no evidence of a material link between the clusters and why it should not be expected. It would be necessary deep wide field CCD photometry for a more conclusive result about the apparent absence of tidal link between the clusters.

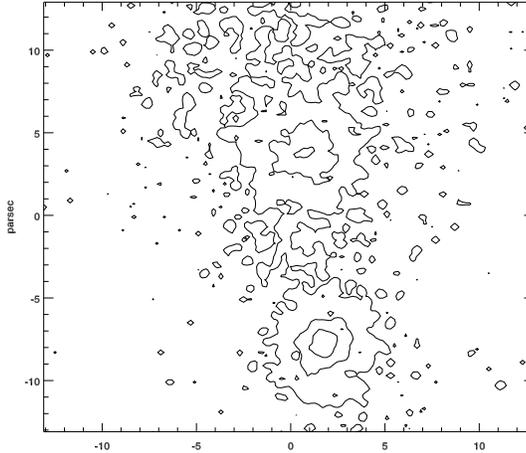

**Fig. 10.** Surface density map for a parabolic encounter of the cluster models NGC 1912 (top) and NGC 1907 (bottom) in the limiting case of a capture, showing the formation of a prominent bridge.

*Acknowledgements.* We made use of information stored at Centre de Données Stellaires (CDS, Strasbourg) and from the DSS (Digitized Sky Survey). We thank the CESUP-UFRGS for alloted time on the CRAY-T94 computer. We are also grateful to an anonymous referee for useful remarks. We acknowledge support from the Brazilian institutions CNPq, CAPES and FINEP. M.R.O. acknowledges financial support from the Brazilian CNPq (no. 200815/00-8), and wishes to thank the LERMA Department of Observatoire de Paris for its hospitality.